\documentclass[aps,amsmath,amssymb,superscriptaddress,notitlepage,groupedaddress,twocolumn,reprint,floatfix]{revtex4-1}
\usepackage{chngcntr}\usepackage{graphicx}
\usepackage{dcolumn}
\usepackage{stackengine}
\usepackage[section]{placeins}
\usepackage[dvipsnames*,svgnames]{xcolor}
\usepackage{verbatim}

\usepackage[position=bottom,caption=false,captionskip=3pt,font=normalsize,subrefformat=simple,labelformat=simple]{subfig}

\usepackage{etoolbox}
\apptocmd{\sloppy}{\hbadness 10000\relax}{}{}

\usepackage{listings}
\usepackage{todonotes}
\usepackage{amsmath}
\usepackage{chngcntr}
\usepackage{bm}
\usepackage{url}

\usepackage{algorithm}
\usepackage{algpseudocode}
\usepackage{xcolor}

\usepackage[breaklinks]{hyperref}
\usepackage{tabularx}
\usepackage{soul}
\usepackage{braket}
\hypersetup{
colorlinks,
linkcolor={blue!100!black},
citecolor={blue},
urlcolor={blue!80!black}}

\begin{document}
\title{Efficient Calculations of the Mode-Resolved \textit{ab-initio} Thermal Conductivity in Nanostructures}
\author{Giuseppe Romano}
\email{romanog@mit.edu}
\affiliation{Institute for Soldier Nanotechnologies, Massachusetts Institute of Technology, 77 Massachusetts Avenue,
Cambridge, Massachusetts 02139, USA}

\begin{abstract}

First-principles calculations of thermal transport in homogeneous materials have reached remarkable predicting power. Modeling deterministically phonon transport in nanostructures, however, poses novel challenges; notably, it entails solving as many algebraic equations as the number of combinations of wave vectors in the discretized Brillouen and polarizations. We show that, within the relaxation time approximation of the Boltzmann transport equation (BTE), this issue is resolved by interpolating the phonon distributions in the vectorial phonon mean free paths (MFP) space. The coupling between structure and mode-resolved heat transport is investigated in terms of angular-resolved bulk thermal conductivity and phonon suppression function, the latter being associated primarily to the material's geometry. Our method, termed the anisotropic MFP-BTE (aMFP-BTE), allows for fast and accurate thermal conductivity calculations in nanomaterials regardless of the number of phonon branches and wave vectors. Furthermore, it naturally blends with first-principles thermal transport calculations, therefore allowing for multiscale, parameter-free simulations. We apply the aMFP-BTE to compute the mode-resolved effective thermal conductivity of porous Si membranes, achieving up to 50x speed with respect to the case with no interpolation. The proposed approach unlocks the engineering of novel nanostructures, with applications to thermoelectrics and heat management. 

\end{abstract}
\maketitle

\section{Introduction}

Tuning thermal transport via nanostructuring is beneficial to several applications, including thermal energy harvesting~\cite{Vineis,lorenzi2018phonon,lee2010effects}, heat management~\cite{kim2007nanostructuring,moore2014emerging} and routing~\cite{anufriev2017heat,zhang2013enhancing}. The key advantage of nanomaterials is the possibility of exploiting ballistic phonon effects, in contrast to macroscopic systems, where transport is mostly diffusive~\cite{chenbook,murthy2005review}. Modeling this transport regime, however, is challenging because it entails solving, at the very least, the Boltzmann Transport Equation (BTE), a much more complicated model than the standard Fourier equation~\cite{Ziman2001}. The steady-state, mode-resolved BTE, in fact, requires tracking phonons in both real and momentum space, and, when solved deterministically, amounts to inverting as many matrices as the number of phonon modes in our system~\cite{murthy1998finite}, which can easily reach hundreds of thousands in realistic materials. To overcome this limitation, several approaches have been proposed; for example in the frequency-dependent BTE (FD-BTE), the BZ is made isotropic from a slice taken along a high-symmetry path, and then sampled in frequency space~\cite{minnich2011quasiballistic,singh2011effect,Loy2013ATransport}. In a previous work, we developed a formalism, called the MFP-BTE, where the sampling is carried out in the MFP-space~\cite{romano2015,romano2016directional}. Recently, a deterministic approach based on the self-adjoint version of the BTE has been proposed~\cite{harter2019prediction}. While these methods initially considered isotropic dispersions, they can be in principle extended to arbitrary anisotropic dispersions, and unnravel the rich physics of the coupling between the crystal structure and nanoscale geometries, as captured by several MonteCarlo simulations~\cite{vermeersch2016cross,wu2016first,Mei2014Full-dispersionNanoribbons,landon2014deviational}. Furthermore, these approaches, unlike single-MFP methods, have allowed for incorporating \textit{first-principle} data~\cite{Romano2014,hao2009frequency,Romano2012}. Building upon the MFP-BTE formalism, we propose a technique to solve efficiently the BTE over the whole BZ, basing on the fact that, in the relaxation-time-approximation (RTA), the nonequilibrium phonon populations are a smooth function of the their vectorial MFPs; we exploit this property by interpolating them onto a regular spherical grid, reducing dramatically computing times while not compromising on the accuracy. Crucially, our method computes heat transport in constant time with respect to the number of phonon branches and wave vectors, opening up the possibility of simulating complex unit-cell materials, such as Bi$_2$Te$_3$ and SnSe- based nanostructures. In deriving our approach, we also introduce the spherical and polar representations of the thermal conductivity and \textit{phonon suppression function}, both pivotal to unraveling the coupling between the geometry and the underlying material. As an example, we first apply our method to Si membrane with infinite thickness, obtaining a speed up of about 50x with respect to the case with no interpolation. A potential speed up of about 3 is found for a three-dimensioanal membrane, which is bound to increase for more complex materials. The proposed method in practice extends the MFP-BTE to anistropic materials, thus we call it the \textit{anisotropic} MFP-BTE, or aMFP-BTE. Taken together, we expect our approach to significantly expands the engineering space of materials for thermoelectrics and heat management applications, while keeping the computational effort amenable to most laptops.

\section{The Mode-Resolved Boltzmann transport equation}

In absence of perturbation, e.g. an applied temperature gradient, the phonon populations are in equilibrium with temperature $T_0$ and are given by the Bose-Einstein distribution
\begin{equation}
    \bar{n}_p(\mathbf{q}) = \frac{1}{e^{\frac{\hbar \omega_p(\mathbf{q})}{k_B T_0}}-1},
\end{equation}
where $p$ is the phonon branch and $\mathbf{q}$ the wave vector. When a temperature gradient is applied, the phonon distributions depart from the equilibrium ones, $n_p(\mathbf{r},\mathbf{q}) = \bar{n}_p(\mathbf{q}) + \Delta n_p(\mathbf{r},\mathbf{q})$, and non-zero current develops. At the steady state, the deviational distributions are obtained by the linearized, time-independent BTE~\cite{Ziman2001},
\begin{eqnarray}\label{bte-mr}
&-&\mathbf{v}_p(\mathbf{q})\cdot\nabla \Delta n_p(\mathbf{r},\mathbf{q}) = \nonumber \\ &=&\sum_{p'} \int_{\mathrm{BZ}} \frac{d\mathbf{q}'}{\Omega_{BZ}}A_{pp'}(\mathbf{q},\mathbf{q}') \Delta n_{p'}(\mathbf{r},\mathbf{q}'),
\end{eqnarray}
where $A_{pp'}(\mathbf{q},\mathbf{q}')$ is the scattering operator and $\mathbf{v}_p(\mathbf{q})$ is the group velocity. Equation~\ref{bte-mr} has six unknowns, three for space and three for momentum. The total number of branches is $N_p$. Let us assume that the BZ has been discretized uniformnly into $N_q$ control volumes $\Delta \mathbf{q}_k$; we can then discretize the momentum space by simply integrating both sides over $\Delta \mathbf{q}_k$ and $\Delta \mathbf{q}_{k'}$. Assuming that distributions and group velocities are constant within the control volume, Eq.~\ref{bte-mr} becomes
\begin{equation}\label{bte}
- \mathbf{v}_\mu \cdot\nabla \Delta n_\mu(\mathbf{r}) = \sum_{\mu'} A_{\mu \mu'}(\mathbf{r}) \Delta n_{\mu'}(\mathbf{r}),
\end{equation}
where $\mu$ collectively indicates phonon branches and wave vectors. We conveniently choose to work in the \textit{temperature formulation} of the BTE~\cite{romano2015,romano2020phonon}
\begin{equation}\label{bte_temp}
    -\mathbf{S}_\mu \cdot \nabla \Delta T_\mu = \sum_\nu W_{\mu\nu} \Delta T_\nu,
\end{equation}
where $\Delta T_\mu = \Delta n_\mu \hbar \omega_\mu /C_\mu$ are the phonon pseudotemperatures (or simply ``temperatures'' hereafter), $W_{\mu \mu'} = A_{\mu \mu'}  \omega _\mu C_{\mu'} C_\mu^{-1} \omega_{\mu'}^{-1}$, and  $\mathbf{S}=\mathbf{v}_\mu C_\mu$; the term $C_\mu = k_B(\eta \sinh{\eta})^{-2}$ is the mode-resolved heat capacity, with $\eta= k_B \hbar \omega_\mu /(2 k_B T_0)$. Within the temperature formulation, the heat flux is $\mathbf{J} = \sum_\mu C_\mu \Delta T_\mu \mathbf{v}_\mu $.

In this work, we employ the relaxation time approximation (RTA), which is accurate for many relevant materials, including silicon~\cite{esfarjani2011heat}. Within RTA, the scattering operator is $W_{\mu\nu}\approx C_\nu \tau_\nu^{-1}\left(\delta_{\mu\nu}\Delta T_\nu - \Delta T^L\right)$, where $\Delta T^L$ is a local pseudotemperature. This quantity is computed by setting $\nabla \cdot \mathbf{J}  = 0$ in Eq.~\ref{bte_temp}, leading to the scattering operator $W_{\mu\nu} = C_\nu/\tau_\nu \left(\delta_{\mu\nu} - a_\nu \right)$, where $a_\nu = C_\nu/\tau_\nu \left[\sum_k C_k/\tau_k \right]^{-1}$. Equation~\ref{bte_temp} then simplifies to
\begin{equation}\label{rta}
  \mathbf{F}_\mu \cdot \nabla \Delta T_\mu^{(n)} + \Delta T_\mu^{(n)} = \sum_\nu a_\nu \Delta T_\nu^{(n-1)},
\end{equation}
where $\mathbf{F}_\mu = \tau_\mu \mathbf{v}_\mu$ is the vectorial MFP. Equation~\ref{rta} is solved iteratively, with $\Delta T_\nu^{(0)}$ given by the diffusive equation~\cite{romano2015}.

We consider periodic materials along both $x$- and $y$- axis, whereas heat flux is enforced by applying difference of temperature $\Delta T_{\mathrm{ext}} = \Delta T_\mu^{\mathrm{R}} - \Delta T_\mu^{\mathrm{L}}$ along the $x$-axis. With $L$ being the distance between the right (R) and left (L) contact, the effective thermal conductivity is computed by
\begin{equation}
    \kappa^{\mathrm{eff}} = -\frac{L}{\Delta T_{\mathrm{ext}} A_{\mathrm{R}}}\int_{\mathrm{R}} \mathbf{J} \cdot \mathbf{\hat{n}} dS,
\end{equation}
where $A_{\mathrm{R}}$ is the area of the right contact. Along the wall we apply the total diffuse scattering boundary condition, e.g. phonons outgoing from a pore are fixed to the boundary temperature~\cite{landon2014deviational}
\begin{equation}\label{TB}
    T_B = \sum_\mu \Delta T_\mu \mathrm{ReLu}(\mathbf{S}_\mu \cdot \mathbf{\hat{n}}) \left[\sum_\nu \mathrm{ReLu}(\mathbf{S}_\nu \cdot \mathbf{\hat{n}}) \right]^{-1},
\end{equation}
where $\mathrm{ReLu}(x)$ is 
\begin{equation}
\mathrm{ReLu}(x) = \begin{cases}
   x,& \text{if } x > 0 \\
    0,              & \text{otherwise}.
\end{cases}
\end{equation}
It's straightforward to show that Eq.~\ref{TB} satisfies the condition $\sum_\mu \mathbf{J}_\mu\cdot \mathbf{\hat{n}} = 0$. Equation~\ref{rta} along with boundary conditions is the mode-resolved BTE under RTA (MR-BTE), and can be discretized using the finite-volume techniques, with $\tau_\mu$, $C_\mu$ and $\mathbf{v}_\mu$ obtained from first-principles calculations. In practice, however, solving directly Eq.~\ref{rta} is challenging: It requires solving an algebraic equation as many times as all the combinations of phonon branch and phonon vectors ($N_p N_q$), which can easily lead to prohibitive computational load. To overcome this issue, we present the \textit{anisotropic} MFP-BTE (aMFP-BTE),  described in the next section.

\section{The anisotropic MFP-BTE}

At each iteration, the mode-resolved RTA-BTE, encoded in Eq.~\ref{rta}, can be computed more efficiently if we exploit the fact that the temperatures are a smooth function in the vectorial MFPs, $\mathbf{F}_\mu$; in fact, we have equations of the form $ f(\mathbf{F},\mathbf{r})  = g(\mathbf{r}) - \mathbf{F}\cdot\nabla f(\mathbf{F},\mathbf{r})$, with $g(\mathbf{r})$ being associated to the local temperature. We can, therefore, solve Eq~\ref{rta} for vectorial MFPs located on a uniform grid and then retrieve $\Delta T_\mu$ by interpolation. We choose a spherical grid with nodes $\mathbf{F}_{mlk}$, where $m$, $l$ and $k$ label the magnitude ($\Lambda_m$), azimuthal ($\theta_l$) and polar angles ($\phi_k$), respectively. The polar and azimuthal angles span uniformly in linear scale, while the magnitudes spread on log scale. 

The generic vectorial MFP is
\begin{eqnarray}
\mathbf{F}_{mkl} =  \Lambda_m\bigg[\sin{\phi_k}\sin{\theta_l}\mathbf{\hat{x}} &+&  \cos{\phi_k}\sin{\theta_l}\mathbf{\hat{y}}+\nonumber \\ &+& \cos{\theta_l}\mathbf{\hat{z}} \bigg] = \Lambda_m \mathbf{S}_{kl},
\end{eqnarray}
where $m$, $l$ and $k$ run up to $N_\Lambda$, $N_\theta$ and $N_\phi$, respectively. The generic mode-resolved phonon temperature is  
\begin{equation}
\Delta T_{\mu} = \sum_{mlk} c_{mlk}^\mu \Delta T_{mlk},
\end{equation}
where $c_{mlk}^\mu$ are linear coefficients such that
\begin{equation}
\mathbf{F}_{\mu} =\sum_{mlk} c_{mlk}^{\mu}  \mathbf{F}_{mlk}.
\end{equation}
Within this formalism, the temperature is
\begin{equation}
 \Delta T^L = \sum_\nu a_\nu \sum_{mlk} a_{mlk}^\mu \Delta T_{mlk};
\end{equation}
finally, Eq.~\ref{rta} becomes
\begin{eqnarray}~\label{mfp}
    \Lambda_m \mathbf{S}_{kl}\cdot\nabla \Delta T_{mkl}^{(n)} &+& \Delta T_{mkl}^{(n)} =\nonumber \\ &=& \sum_{m'k'l'} a_{m'k'l'} \Delta T_{m'k'l'}^{(n-1)},
\end{eqnarray}
where $a_{mkl}=\sum_{u} c_{mkl}^{\mu}$. Similarly, heat flux is given by $\mathbf{J} = \sum_{mkl} \Delta T_{mkl}  \mathbf{G}_{mkl}$, where $\mathbf{G}_{mkl}=\sum_\mu c_{mkl}^{\mu} C_\mu \tau_\mu^{-1}\mathbf{F}_\mu$. Upon convergence, the effective thermal conductivity is now provided in terms of $\Delta T_{mkl}$, i.e.
\begin{equation}\label{kappa_mode}
    \kappa_{\mathrm{eff}} = \sum_\mu \frac{C_\mu}{\tau_\mu} F_{\mu,x} \langle \Delta T_\mu \rangle = \sum_{mkl}  G_{mkl,x}\langle \Delta T_{mkl} \rangle,
\end{equation}
where $\langle f \rangle =  -L \Delta T_{\mathrm{ext}}^{-1} A_{\mathrm{R}}^{-1} \int_\mathrm{R} f  dS $.
To better understand the effect of the geometry on $\kappa_{\mathrm{eff}}$, it is convenient to define a suppression function, which is a measure on how much heat is carried in the nanomaterial compared to that from the bulk. Originally, this tool was conceived as a MFP- or frequency-dependent function~\cite{minnich2012determining} and later was generalized to include directionality~\cite{romano2016directional}. Here we define the \textit{mode-resolved} suppression function as  $S_\mu = \langle \Delta T_\mu \rangle \left(v_{\mu, x}\tau_\mu\right)^{-1}$, which is calculated with respect to the mode-specific vectorial MFP projected onto the direction of the applied temperature. Using $S_\mu$, the effective thermal conductivity is $\kappa^{\mathrm{eff}} = \sum_\mu \kappa_\mu^{xx} S_\mu = \sum_{mkl} \kappa_{mkl} S_{mkl}$, where 
\begin{equation}\label{kappa}
\kappa_{mkl} = \sum_\mu \frac{C_\mu}{\tau_\mu }F_{\mu,x} a_{mkl}^\mu F_{mkl,x} 
\end{equation}
\begin{equation}\label{sup}
S_{mkl} = \frac{\langle \Delta T_{mkl}\rangle}{F_{mkl,x}},
\end{equation}
are defined as the spherical bulk thermal conductivity and spherical suppression function, respectively. Equation~\ref{mfp} along with the definitions~\ref{kappa}-\ref{sup} constitute the aMFP-BTE model. The corresponding workflow is summerized in algorithm~\ref{algo}.

\begin{algorithm}[H]
    \caption{aMFP-BTE. Note that the $mkl$ labels have been grouped into $g$.}
    \label{algo}
    \begin{algorithmic}[1] 
            \State Solve $ \Delta T^{L,(0)} \gets \nabla^2\Delta T^{L,(0)} = 0$ 
             \State $n = 1$
            \While{error $>1e^{-3}$} 
                \State $\Delta T^{L,(n+1)} = 0$
                \State $\kappa^{\mathrm{eff},(n+1)} = 0$
                \For{g = 1:$N_\Lambda N_{\theta} N_{\phi}$} 
                  \State $  \Delta T_g^{(n+1)} \gets \mathbf{F}_g\cdot\nabla \Delta T_g^{(n+1)}  + \Delta T_g^{(n+1)} = \Delta T^{(0)} $
                  \State $  \Delta T^{L,(n+1)} \gets T^{L,(n+1)} +   a_g \Delta T_g^{(n+1)}$
                  \State $S_g  \gets \langle\Delta T_g^{(n+1)} \rangle F_{g,x}^{-1}$
                  \State $\kappa^{\mathrm{eff},(n+1)}  \gets  \kappa^{\mathrm{eff},(n+1)}  + \kappa_g  S_g$
                \EndFor   
                \State error $\gets |(\kappa^{\mathrm{eff},(n+1)} - \kappa^{\mathrm{eff},(n)})|/\kappa^{\mathrm{eff},(n+1)}$ 
                \State $n \gets n +1$
            \EndWhile\label{euclidendwhile}
    \end{algorithmic}
\end{algorithm}

\section{Membranes with Infinite Thickness}

We first apply the aMFP-BTE to porous Si membrane with circular pores and infinite thickness. The simulation domain comprises a square unit cell of size $L$, containing one circular pore, to which a difference of temperature $\Delta T_{\mathrm{ext}}$ is applied. The chosen porosity is 0.2. The scattering times at 300 K, group velocities and phonon frequencies are computed with density functional theory and supercell approach, implemented in AlmaBTE~\cite{Carrete2017AlmaBTEMaterials}, which also provides pre-computed second- and third- order force constants. Convergence was found with a uniform wave vector grid of 32x32x32 points, to which it corresponds a bulk thermal conductivity, $\kappa_{\mathrm{bulk}}$, of~160 Wm$^{-1}$K$^{-1}$. Naturally occurring isotope disorder was included in the calculation. At the macroscopic level, reduction in thermal transport is given by Fourier's law, which, in our case, gives ~$\kappa_f$=107.4 W m$^{-1}$K$^{-1}$, in line with the prediction from Eucken theory~\cite{hasselman1987effective}, i.e. $\kappa_f \approx \kappa_{\mathrm{bulk}}\frac{1-\phi}{1+\phi}$=106.7 W m$^{-1}$K$^{-1}$. 
\begin{figure}[htp!]
\includegraphics[width=0.48\textwidth]{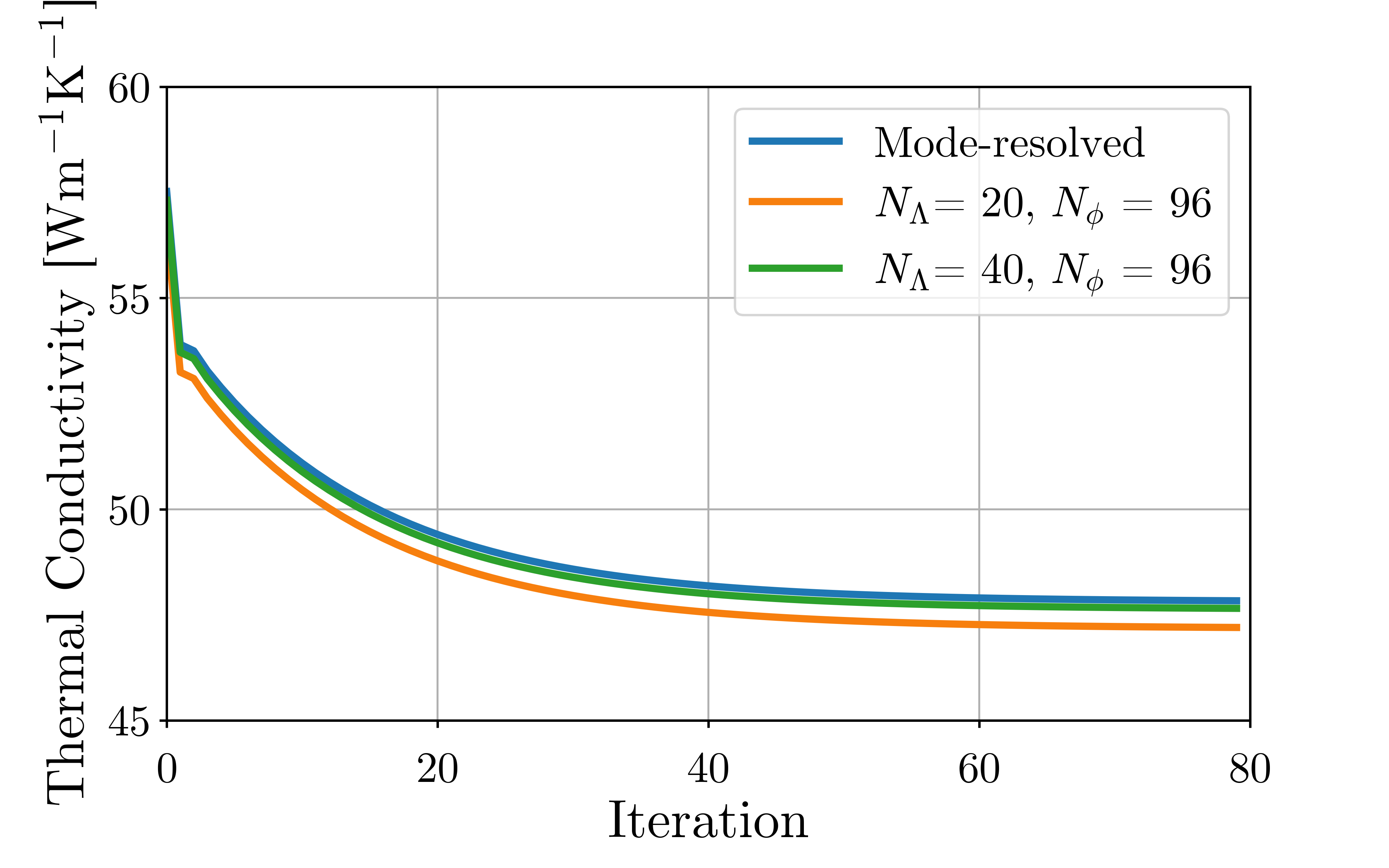}
\caption{The effective thermal conductivity versus the number of iterations for the case with L = 200 nm. Results are shown for the MR-BTE and the aMFP-BTE, for different spherical discretizations. In this work, we choose $N_\Lambda$ = 40 and $N_\phi$ = 96. \label{fig1}\label{convergence}}
\end{figure}
The value for $\kappa_f$ does not vary with $L$, as long as the porosity is kept constant. On the other side, phonon size effects, computed by the BTE, depend on the size of the unit cell, as shown below.
\begin{figure*}[htp!]
\subfloat[\label{fig1a}]
{\includegraphics[width=0.48\textwidth]{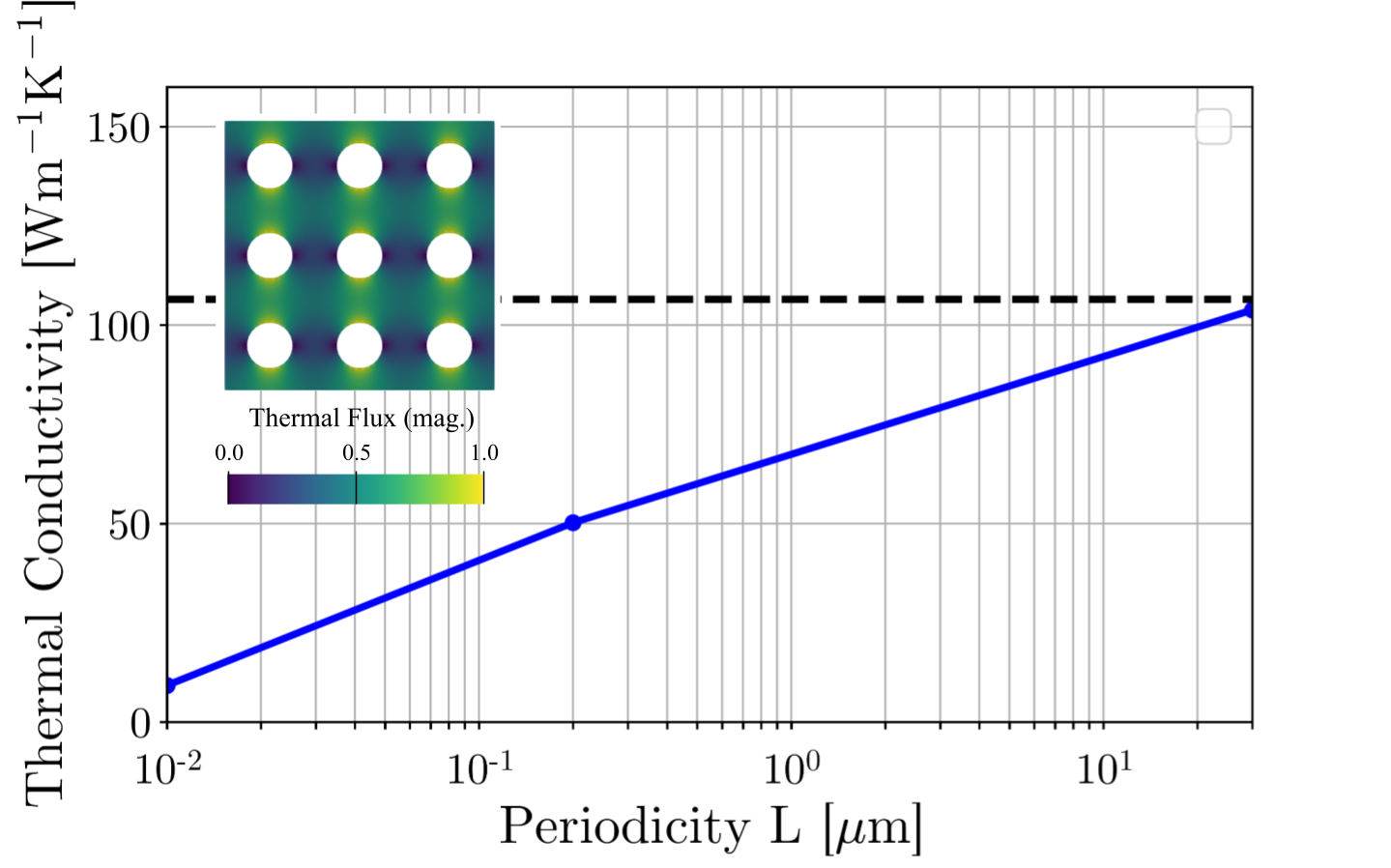}}
\hfill
\subfloat[\label{fig1b}]
{\includegraphics[width=0.48\textwidth]{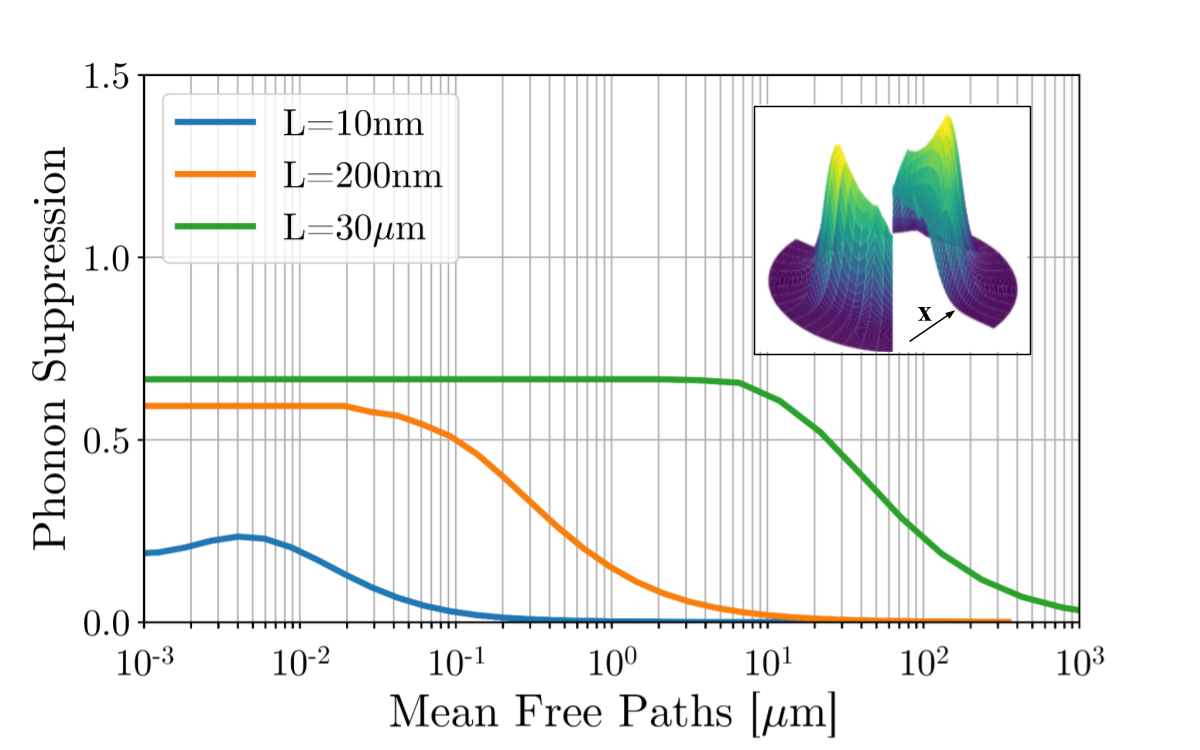}}
\quad
\subfloat[\label{fig1c}]
{\includegraphics[width=0.48\textwidth]{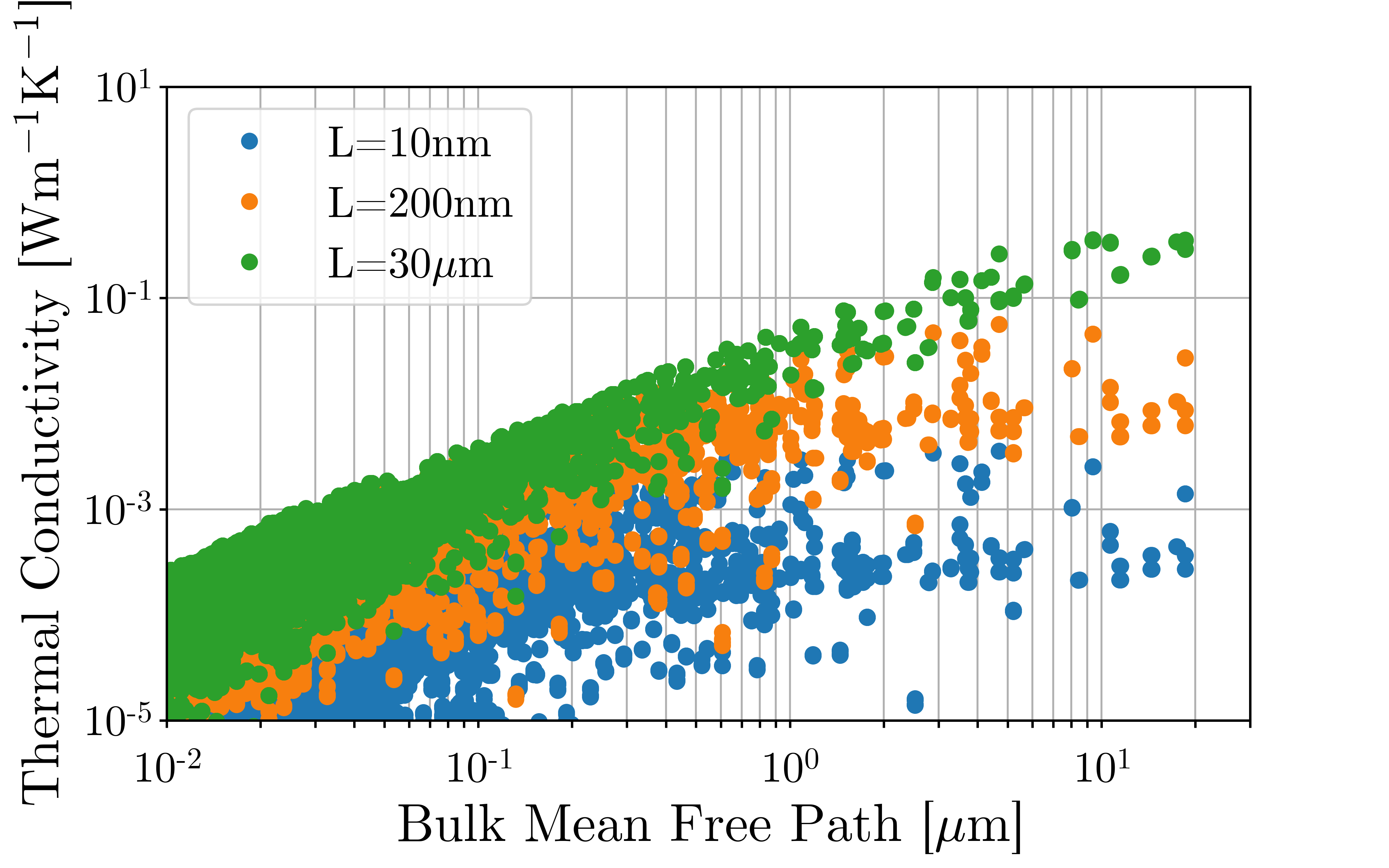}}
\hfill
\subfloat[\label{fig1d}]
{\includegraphics[width=0.48\textwidth]{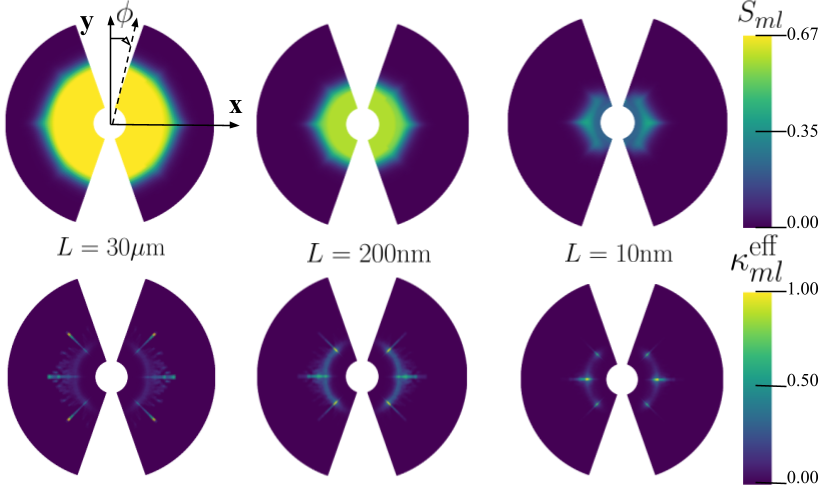}}
\caption[]{a) The effective thermal conductivity of porous Si with porosity 0.2 and periodicity L = 10 nm, 200 nm and 30 $\mu$m. In the inset, a map of the magnitude of thermal flux (normalized to 1) b) The average phonon suppression function for different periodicities. In the inset, a 3D represention of $S_{tk}$ for the case with L = 10 nm. c) Mode-resolved $\kappa^{\mathrm{eff}}_\mu$ versus the projected MFP $\mathbf{F}_\mu \cdot \mathbf{\hat{x}}$ for different periodicities. d) Upper panel: polar phonon suppression function, $S_{tk}$. A difference of temperature is imposed along the $x$-axis. Bottom panel: polar effective thermal conductivity, $\kappa_\mu^{\mathrm{eff}}$.}
\end{figure*}
In the case with infinite thickness, the BTE is a two-dimensional system but still with the azimuthal angle explicitly taken into account, i.e.
\begin{eqnarray}
&\Lambda_m&  \sin(\theta_l) \bigg[\sin(\phi_k) \frac{\partial \Delta T_{mkl} }{\partial x} + \nonumber \\ &+& \cos(\phi_k) \frac{\partial \Delta T_{mkl} }{\partial y}\bigg] +\Delta T_{mkl} = \Delta T_L.
\end{eqnarray}
Unlike with the 3D case, the phonon temperatures are a smooth function of $\Lambda_m \sin(\theta_l)$, which then can be reformulated as an ``effective'' MFP $\xi_t$. Leveraging this result, the interpolation is performed between the bulk MFPs projected onto the \textit{xy} plane, and
\begin{equation}
\mathbf{F}_{tk} = \xi_t\left[ \sin{(\phi_k)}\mathbf{\hat{x}} + \cos{(\phi_k)}\mathbf{\hat{y}} \right],
\end{equation}
with the corresponding interpolation coefficients being $c_{tk}^{\mu}$. The polar representations of the bulk thermal conductivity and suppression function are $\kappa_{tk} = \sum_\mu C_\mu/\tau_\mu F_{\mu,x} c_{tk}^\mu F_{\mu,x}$ and $S_{mt}=\langle \Delta T_{tk}\rangle/F_{tk,x}$, respectively. Since only the discretization in the polar angle and MFPs is needed, this treatment greatly enhances the computational efficiency, unlocking realistic phonon transport simulations on common laptops.

To assess the efficiency of the proposed model, we solve the MR-BTE, i.e. Eq.~\ref{bte_temp}, and the aMFP-BTE, defined by Eq.~\ref{mfp}. The mode resolution is the same as the q-grid used for bulk thermal conductivity calculations, $N_pN_q$ = 196608. On the other hand, we choose a variable spherical grid for the aMFP-BTE, obtaining an agreement between the two models within $0.5\% $ with $N_\Lambda$ = 40 and $N_\phi$ = 96, as shown in Fig.~\ref{convergence}. We note that for the aMFP-BTE case, the ``interpolated modes'' were only 3840, obtaining roughly a 50x speed up with respect to the MR-BTE. 

Figure~\ref{fig1a} shows $\kappa^{\mathrm{eff}}$ for L = 50 nm, 200 nm and 30 $\mu$m. The last case employs a multiscale approach, which will be discussed elsewhere. As expected, strong suppressions are achieved for smaller periodicities since the distance between the pores' walls shrinks and phonon transport is suppressed over a wider range of the MFP distribution~\cite{romano2015}. For large L, we approach the diffusive limit, described above. As shown in the inset of~\ref{fig1a}, thermal flux is concentrated in the space between the pores, a well-known signature of ballistic transport~\cite{anufriev2020ray}.

The influence of the structure on thermal transport is best described by the suppression function, as detailed in the previous section. Its angular average, given by $S_t = \sum_k S_{tk}$ and illustrated in Fig.~\ref{fig1b}, shows three regimes~\cite{minnich2012determining}: (i) the large-MFP regime, namely where transport is mostly ballistic, $S_t\approx \xi_t^{-1}$; (ii) the intermediate-MFP regime, where both ballistic and diffusive transport are present and (iii) the small-MFP regime, where transport is mainly diffusive. Although all the configurations have the same porosity, their small-MFP limits do not match the prediction from Fourier's law. In fact, in nongray materials, ballistic phonons may also effect those who travel diffusively via the definition of $\Delta T^L$~\cite{romano2019diffusive}. For structures whose characteristic length is larger than most heat-carrying phonons, however, $S_0 \approx \kappa_{\mathrm{fourier}}/\kappa_{\mathrm{bulk}}$. Once $\Delta T_{tk}$ are computed, it is possible to compute the mode-resolved temperature $\Delta T_\mu = \sum_{tk} c_{tk}^{\mu} \Delta T_{tk}$; from the mode-resolved temperature, we can then obtain $S_\mu$ and thus the mode resolved effective thermal conductivity $\kappa^{\mathrm{eff}}_\mu = C_\mu \mathbf{v}_\mu F_{\mu,x}  S_\mu$. In Fig.~\ref{fig1c}, we plot $\kappa^{\mathrm{eff}}_\mu$ versus $\mathbf{F}_{\mu,x}$ for different L; we note that, as L becomes smaller, a wider spectrum of the MFP is suppressed. 

The aMFP-BTE allows to explore the coupling between the structure and the material at the angular level. In the case with infinite thickness, only polar discretization is needed thus we can plot $S_{tk}$ as a surface in 3D, as shown in the inset of Fig.~\ref{fig1b} for the case with L = 50 nm; we note there are two main lobes, corresponding to the forward and backward direct paths~\cite{romano2016directional}. In Fig.~\ref{fig1d}, we show the top-view of the polar suppression function for all L as well the polar representation of $\kappa^{\mathrm{eff}}$. For L = 10 nm, $S_{tk}$ is strongly anisotropic showing four additional lobes. For L = 200, however, this anisotropy becomes less pronounced, and for L = 30 $\mu$m, $S_{tk}$ is mostly isotropic. This last case can be regarded as the diffusive limit. Note that for $\mathbf{F}_{tk}$ such that $|\mathbf{F}_{tk} \cdot \mathbf{\hat{y}}|/|\mathbf{F}_{tk}|$ is small, numerically instabilities regarding the calculation of $S_{tk}$ occur. To understand this issue, it is convenient to rewrite the temperatures as $\Delta T_{tk} = \Delta T^L - \mathbf{F}_{tk}\cdot\nabla T_{tk}$. The suppression function then reads
\begin{equation}\label{sup2}
    S_{tk} = -\frac{\partial \langle \Delta T_{tk} \rangle }{\partial x} - \frac{1}{\tan{\phi_k}}\frac{\partial \langle \Delta T_{tk}\rangle }{\partial y};
\end{equation}
the second term of Eq.~\ref{sup2}, for $\phi_k\approx 0$ and $\phi_k\approx \pi$, becomes numerically challenging. For this reason, in Fig.~\ref{fig1c}, we plot $S_{tk}$ only for regions far away from those two critical cones. We note that, however, these inaccuracies do not alter $\kappa^{\mathrm{eff}}$ since  $\kappa^{\mathrm{bulk}}_\mu\approx 0$ for $\mathbf{F}_\mu$ aligned with $\mathbf{\hat{y}}$ and -$\mathbf{\hat{y}}$. Furthermore, $S_{tk}$ is computed only for analysis purposes, while in the actual calculation for $\kappa^{\mathrm{eff}}$ we use Eq.~\ref{kappa_mode}, which overcomes these inaccuracies.

Similarly to $S_{tk}$, we explore the polar representation of the effective thermal conductivity, i.e. $\kappa^{\mathrm{eff}}_{tk} = \kappa^{\mathrm{bulk}}_{tk} S_{tk}$, shown in Fig.~\ref{fig1d}. For all L, we note that most heat is carried by phonons along high-symmetry axis, and that the anisotropy of $\kappa^{\mathrm{eff}}_{tk}$ is influenced by $L$. In fact, for the case with L = 200 nm, we can also observe a circular band in addition to peaks around high-symmetry axis. Furthermore, we note that the case with L = 10 nm, the extra lobes in $S_{tk}$ are aligned with the 100 axis of the crystal, thus capturing heat transport in that region of the polar space. Lastly, for L = 30 $\mu$m, $\kappa^{\mathrm{eff}}_{tk}\approx \kappa_{tk}$; this \textit{filtering} effect is made accessible thanks to the aMFP-BTE and can be exploited to enhance thermal transport tunability.

\section{Membranes with Finite Thickness}\label{finite}

\begin{figure}[htp]
\includegraphics[width=0.48\textwidth]{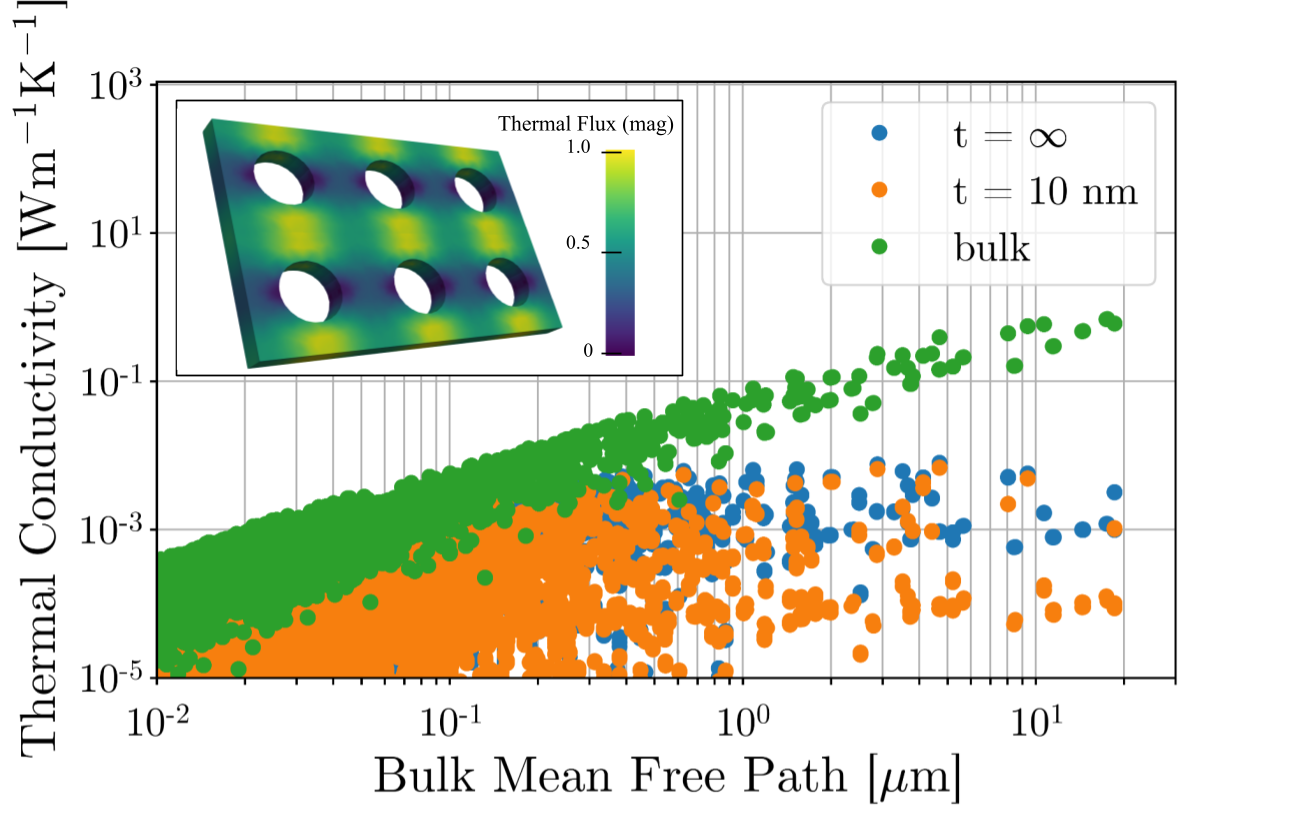}
\caption{Mode-resolved effective thermal conductivity, $\kappa_\mu^{\mathrm{eff}}$, versus the projected MFP $\mathbf{F}_\mu \cdot \mathbf{\hat{x}}$, for bulk and for thicknesses t = 10 nm and ~$\infty$. In the inset, a 3D map of the magnitude of thermal flux, normalized to 1.\label{fig2}}
\end{figure}

For membranes with finite thickness, thermal transport is further reduced due to the top and bottom surfaces scattering phonons. In this case, the characteristic size is determined by both pore-pore distance and thickness. To show this effect, we consider a periodic membrane with circular pores, porosity 0.2, $L$ = 50 nm and thickness t = 10 nm. Convergence was found with $N_\Lambda$=30, $N_\phi$ = 48 and $N_\theta$ = 48, at which it corresponds $\kappa^{\mathrm{eff}}\approx$  11.3 W m$^{-1}$K$^{-1}$. The case with infinite thickness (t=$\infty$), modeled as described in the previous section, gives ~23.7W m$^{-1}$K$^{-1}$. The mode-resolved $\kappa^{\mathrm{eff}}$, reported in Fig.~\ref{fig2}, reveals that most modes have stronger suppression with respect to the case with t=$\infty$; these modes are those who scatter with the top and bottom surfaces. On the other side, there still a small fraction of phonons that travel without being strongly influenced by the top and bottom surfaces, i.e. those with vectorial MFPs that are aligned with the $x$-axis. In the inset of Fig.~\ref{fig2}, we report a 3D map of the magnitude of thermal flux, which, analogously to the case with $t=\infty$, shows high values in the space between pores. 

Lastly, we note that in this case the speed gain with respect to the MR-BTE is around 3, assuming that convergences occur with comparable number of steps. However, with more complex materials, the number of branches as well as the number of wave vectors increases, with little or no increase in the computational demand from aMFP-BTE; therefore, the aMFP-BTE may introduce large speed ups for 3D systems, as well.

\section{Conclusion}

We introduce the aMFP-BTE, a method that reduces dramatically the computational demand of thermal conductivity calculations in nanostructures, leveraging interpolation of phonon distributions in the vectorial MFP space. The application of the aMFP-BTE to Si porous membranes with infinite thickness reveals a speed-up of about 50x. Taking into account the whole BZ, our method is able to elucidate on the interplay between the anisotropy of the bulk thermal conductivity and the material's geometry, highlighting novel mechanisms for thermal transport tuning. The aMFP-BTE solves for thermal transport in constant time with respect to the number of phonon branches and wave vectors, opening up the possibility of simulating phonon size effects in promising thermoelectric materials, such as SnSe and Bi$_2$Te$_3$. The code used for this work will be released with the package OpenBTE~\cite{romano2019openbte}.

\begin{acknowledgments} 
Research was partially supported by the Solid-State SolarThermal Energy Conversion Center (S3TEC), an Energy
Frontier Research Center funded by the U.S. Department of
Energy (DOE), Office of Science, Basic Energy Sciences
(BES), under Award No. DESC0001. The author thanks Steven G. Johnson for helpful discussions.
\end{acknowledgments}

\appendix

\section{Finite-Volume Formulation of the BTE}

We discretize Eqs.~\ref{bte-mr}-\ref{mfp} using the finite-volume techniques. For simplicity, we label the phonon temperatures with $\mu$, but the same procedure holds for both the MR-BTE and the aMFP-BTE. The mesh, generated by GMSH~\cite{geuzaine2009gmsh}, is unstructured, with elements at the border of domain matching their periodic counterparts. The BTE for a generic $\mathbf{F}_\mu$ reads
\begin{equation}\label{deltat}
    \mathbf{F}_\mu\cdot \nabla \Delta T_\mu(\mathbf{r}) + \Delta T_\mu(\mathbf{r}) = \Delta T^L(\mathbf{r}),
\end{equation}
which, after integrating both side over the control volume $V_c$, becomes
\begin{equation}\label{deltat2}
    \frac{1}{V_c}\int_{\partial V_c}\Delta T_\mu\mathbf{F}\cdot\mathbf{\hat{n}} dS + \Delta T_{c\mu} = \Delta T^L_c,
\end{equation}
where we uses Gauss' theorem. The RHS of Eq.~\ref{deltat2} is simply given by $\Delta T^{L}_c = \sum_\nu a_\nu \Delta T_{c\nu}$. The term associated to the surface integral is discretized according to the upwind method~\cite{murthy1998finite,romano2011multiscale}, yielding
\begin{equation}\label{rcc}
 \frac{1}{V_c}\int_{\partial V_c}\Delta T\mathbf{F}_\mu \cdot\mathbf{\hat{n}} dS = I_{c\mu}^{\mathrm{in}} + I_{c\mu}^{\mathrm{out}}+ I_{c\mu}^{\mathrm{B}} + I_{c\mu}^{\mathrm{P}}.
\end{equation}
The contribution $I_{c\mu}^{\mathrm{in}}$ accounts for all the flux incoming to a given element, and is given by
\begin{equation}
I_{c\mu}^{\mathrm{in}} = -\sum_{c'}  \mathrm{ReLu}(-\mathbf{F}_\mu\cdot \mathbf{R}_{cc'} ) \Delta T_{c'\mu},
\end{equation}
where
\begin{equation}
\mathbf{R}_{cc'} =  \begin{cases}
   \frac{1}{V_c}A_{cc'}\mathbf{\hat{n}}_{cc'}',& \text{if } c \text{ and } c' \text{ are neighbors}\\
    0,              & \text{otherwise}.
\end{cases}
\end{equation}
The terms $\mathbf{\hat{n}}_{cc'}$ and $A_{cc'}$ are the normal (pointing toward the volume $c'$) and the area of the side between the volume $c$ and $c'$. The ReLu function filters only the incoming contributions to the element $c$, while $\mathbf{R}_{cc'}$ represents the connections between the volumes.
The contribution $I_{c\mu}^{\mathrm{out}}$ is the flux leaving the volume, and reads
\begin{equation}
I_{c\mu}^{\mathrm{out}} = \sum_{c''}  \mathrm{ReLu}(\mathbf{F}_\mu\cdot \mathbf{R}_{cc''} ) \Delta T_{c''}.
\end{equation}
In the upwind scheme, this term amounts to the diagonal of the stiffness matrix. The third term of Eq.~\ref{rcc} arises from the flux bouncing back from an adiabatic boundary, and is given by
\begin{equation}
I_{c\mu}^{\mathrm{B}} = -\sum_{s}  \mathrm{ReLu}(-\mathbf{F}_\mu\cdot \mathbf{\hat{n}}_{s} ) g_{sc} \Delta T_s.
\end{equation}
where $\mathbf{\hat{n}_s}$ is the normal of the surface pointing outward with respect to the computational domain, and $T_s$ is the boundary temperature, computed with Eq.~\ref{TB}. The term $g_{sc}$ is
\begin{equation}
g_{sc} =  \begin{cases}
   \frac{1}{V_{c}}A_s ,& \text{if } s \text{ is a side of } c\\
    0,              & \text{otherwise},
\end{cases}
\end{equation}
where $A_s$ is the area of the side $s$. Using Eq.~\ref{TB}, we can rewrite the boundary contribution as $I_{c\mu}^{\mathrm{B}}=\sum_\nu H_{\mu\nu c} \Delta T_{c\nu}$, where
\begin{equation}
 H_{c\mu\nu} = -\sum_{s} \mathrm{ReLu}(-\mathbf{F}_\mu\cdot \mathbf{\hat{n}}_{s} ) g_{sc} \frac{\mathrm{ReLu}(\mathbf{S}_\nu\cdot\mathbf{\hat{n}}_s) }{\sum_k\mathrm{ReLu}(\mathbf{S}_k\cdot\mathbf{\hat{n}}_s)}.
\end{equation}
 The fourth and last contribution is due to the periodic boundary conditions, 
\begin{equation}
I_{c\nu}^{\mathrm{P}} = -\sum_{c'}  \mathrm{ReLu}(-\mathbf{F}_\nu\cdot \mathbf{R}_{cc'}^P ) \Delta T_{\mathrm{ext}}
\end{equation}
where
\begin{equation}
\mathbf{R}_{cc'}^P =  \begin{cases}
   \frac{1}{V_c}A_{cc'}\mathbf{\hat{n}}_{cc'},& \text{if } c \text{ and } c' \text{ are periodic}\\
    0,              & \text{otherwise}.
\end{cases}
\end{equation}
Putting these contributions together, along with Eq.~\ref{deltat}, we obtain the iterative linear system
\begin{equation}\label{iter5}
    \sum_{c'} A_{cc'\mu} \Delta T_{c'\mu}^{(n)} = \sum_\nu (a_\nu +  H_{c\mu\nu})  \Delta T_{c\nu}^{(n-1)}  + I_{c\mu}^P 
\end{equation}
where
\begin{eqnarray}
    A_{cc'\mu} &= \delta_{cc'}  + \sum_{c''} \mathrm{ReLu}(\mathbf{F}_\mu\cdot \mathbf{R}_{cc''} )  \delta_{cc'} - \nonumber \\ &-\mathrm{ReLu}(-\mathbf{F}_\mu\cdot \mathbf{R}_{cc''} ) \delta_{c'c''} .
\end{eqnarray}
Finally, we can write Eq.~\ref{iter5} in vector notation
\begin{equation}
    \mathbf{A}_\mu \mathbf{\Delta T_\mu}^{(n)}=\sum_\nu \mathbf{S}_{\mu\nu} \mathbf{\Delta T}_\nu^{(n-1)} + \mathbf{I}_\mu,
\end{equation}
where $S_{c\mu\nu} = \alpha_\nu + H_{c\mu\nu}$.

\bibliography{biblio} 

\end{document}